# Vector erf-Gaussian beams: fractional optical vortices and asymmetric TE and TM modes

T. Fadeyeva[1], C. Alexeyev[1], A. Rubass[1] and A. Volyar[1,*]

[1]Department of Physics, Taurida National V.I. Vernadsky University, Vernadsky av.4, Simferopol, Ukraine, 95007

*Corresponding author: volyar@crimea.edu

We have considered the paraxial vector erf-Gaussian beams with field distribution in the form of the error function that are shaped by the cone of plane waves with a fractional step of the azimuthal phase distribution modulated by the Gaussian envelope. We have revealed that the initial distributions of the transverse electric and transverse magnetic fields have the form far from standard ones but at the far diffraction field the field distributions recover nearly the symmetric form. We have also revealed that the half-charged vortices in one of the field components can propagates up to the Rayleigh length without essential structural transformations but then splits into an asymmetric net of singly charged vortices.
*OCIS codes:260.0260,350.5030,260.6042*

The ability of the cylindrically polarized beams to be focused into a subwave-sized spot [1] initiates a great deal of technical applications for manipulating microparticles, shaping special structures on the solid surfaces etc (see e.g. [2,3]). However, there is a little restriction – the dark spot of the wave beam where amplitudes of the beam components vanish, is inevitably positioned at the beam axis. Such a limitation is an intrinsic feature of the cylindrically polarized beams (transverse electric TE and transverse magnetic TM ones) and can scarcely be overcome by traditional ways. Indeed their most simple model is of the conical fantail of plane waves with typical phase distributions: $\phi = -\varphi$ in the right hand polarized (RHP) component and $\phi = \varphi$ in the left hand polarized (LHP) one [4,5] ($\varphi$ stands for the azimuthal angle in the polar coordinates). Symmetry of the phase distribution $\phi$ in the wave fantail and destructive interference ensure shaping the centered optical vortices in each polarized component. The phase symmetry can be disturbed by inserting the fractional optical vortices in the beam components so that, say, the phase dependence of the RHP component is $\phi = p\varphi$, where $p$ is a fractional number. However, Berry have shown [6] that such a beam with Gaussian envelope splits into the infinite series of the vortex-beams with integer topological charges. Further theoretical and experimental analysis (see, e.g. [7-9] and references therein) have demonstrated a radical transformation of the axial symmetry of the beams at very short beam lengths.

At the same time, there are at least two key questions in the problem. The first, can the beams with fractional optical vortices propagate at the distances comparable with the Rayleigh length $z_0$ without radical structural transformations? The second, can the TE and TM vector structures exist in such singular beams? It is these questions that are considered in our paper.

The aim of the paper is to analyze the vector structure of the beam fields with circularly polarized components bearing optical vortices with fractional topological charges.

1. It is well-known (see, e.g. [10]) that solutions of the Maxwell equations for monochromatic wave beams in a homogeneous medium can be reduced to the vector Helmholtz equation for the vector potential **A** under the condition of Lorentz gauge so that the electric **E** and magnetic **H** fields in free space are written as

$$\mathbf{H} = \nabla \times \mathbf{A}, \quad \mathbf{E} = i k \mathbf{A} + i\nabla(\nabla \mathbf{A})/k, \quad (1)$$

where $k$ is a wavenumber. In the paraxial approximation $|\partial_z^2 \mathbf{A}| \ll |k^2 \mathbf{A}|$ the longitudinal components of the electric and magnetic fields can be written out in terms of the transverse field components $\mathbf{E}_\perp$, $\mathbf{H}_\perp$ as $E_z \approx i\nabla_\perp \mathbf{E}_\perp / k$, $H_z \approx i\nabla_\perp \mathbf{H}_\perp / k$, where $\nabla_\perp = \mathbf{e}_x \partial_x + \mathbf{e}_y \partial_y$, while the complex amplitude $\tilde{\mathbf{A}}$ of the vector potential $\mathbf{A} = \tilde{\mathbf{A}}(x,y,z)\exp(ikz)$ obeys the vector paraxial equation $(\nabla_\perp^2 + 2ik\partial_z)\tilde{\mathbf{A}}_\perp = 0$. If we take **A** to be directed along the x-axis: $\mathbf{A} = \mathbf{e}_x \tilde{A}_\perp \exp(ikz) = \mathbf{e}_x \Psi(x,y,z)\exp(ikz)$, then the solution to the paraxial wave equation for the scalar function $\Psi$ can be presented in the form [11]:

$$\Psi = \exp(-K^2/(2ik Z))F(X,Y)G(x,y,z), \quad (2)$$

where $Z = z - iz_0$, $X = x/Z, Y = y/Z$, $G = \exp\{ikr^2/(2Z)\}/Z$ is the Gaussian envelope, $z_0 = kw_0^2/2$, $w_0$ stands for the radius of the beam waist, $K$ is the additional beam parameter (including complex values) while the function $F(X,Y)$ satisfies the 2D Helmholtz equation:

$$(\partial^2/\partial X^2 + \partial^2/\partial Y^2 + K^2)F = 0. \quad (3)$$

The function $F$ obeying eq. (3) can be written in the cylindrical coordinates $(R,\varphi)$ as

$$F(R,\varphi) = \int_0^{2\pi} \exp\{i[p\varphi' + K R\cos(\varphi' - \varphi)]\}d\varphi', \quad (4)$$

where $X = R\cos\varphi, Y = R\sin\varphi$, $R^2 = r^2/Z^2$, $r^2 = x^2 + y^2$. In case the index $p$ is integer, the eq. (2) describes the standard Bessel-Gaussian beams [11]. In our consideration we will treated the number $p$ as a fractional one, say, $p = -1/2$. In our model it means that the conical bundle of plane waves with a step phase $\phi = p\varphi$ is modulated by the radial Gaussian envelope in the **k**-space. Thus, the electric field of the paraxial beam can be written as $E_x \approx ik\Psi$, $E_y \approx 0$, $E_z \approx -\partial_x \Psi$. By using eqs (4) and (2) we obtain the explicit form of the beam for $p = -1/2$:

$$\Psi = e^{-i\varphi/2 + iK^2/(2kZ)} \left\{ e^{\Re^2/2} erf(\Re_s) - e^{-\Re^2/2} erf(i\Re_c) \right\} G/\Re, \quad (5)$$

where $\Re_s = \Re \sin(\varphi/2)$, $\Re_c = \Re \cos(\varphi/2)$, $\Re = \sqrt{2iKR}$, $erf(x)$ is the error function. In what follows we will call later on such a beam – *the erf-Gaussian beam (erf-G)*.

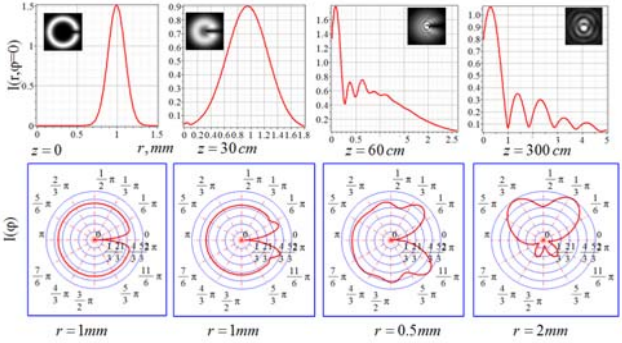

Fig.1. (Color online) Evolution of the intensity distributions $I(r, \varphi = 0, z = const)$ and $I(r = const, \varphi, z = const)$ in the erf-G beam with $K = 10^4$, $w_0 = 150 \mu m$

There are two different types of the erf-G beams depending on the choice of the parameter $K$: 1) $K$ is a purely real value; 2) $K$ is a purely imaginary value. Fig. 1 illustrates the evolution of the intensity distribution in the erf-G beam of the first type with $K = 10^4$ and the beam waist $w_0 = 150 \mu m$. At the initial plane z=0 the beam intensity has a C-shaped form with a vanishingly small intensity along the ray $\varphi = 0$ so that the field has a form $\Psi \sim \sin(\varphi/2)\exp(-i\varphi/2)$ near the axis. It should be noted that Soskin et al [9, 13] and Kiselev (see the comment 17 in [11]) have denoted facilities of existing of similar field structures. The phase function $\Phi(r,\varphi)$ of the beam shown in Fig.2 has a helicoid-like form with the phase step equal to $-\pi$ that corresponds to the half-integer optical vortex. As the beam propagates along z-axis, its form is gradually reconstructed (see Fig.1), the fractional step of the phase is gradually smoothed at the length $z \approx 10 cm \sim z_0$ while the net of singly-charged optical vortices are nucleated around the axis.

Absolutely different situation is observed for the second type of the erf-G beams with the imaginary $K$ - parameter shown in Fig.3. There is an asymmetric net of optical vortices at the initial plane z=0. The beam propagation is accompanied by the structural reconstruction of the field distribution. The oppositely charged vortices in the net rapidly annihilate at the distance about $z = 3cm$ and the vortex-free field in the mirror-imaged C-form is shaped.

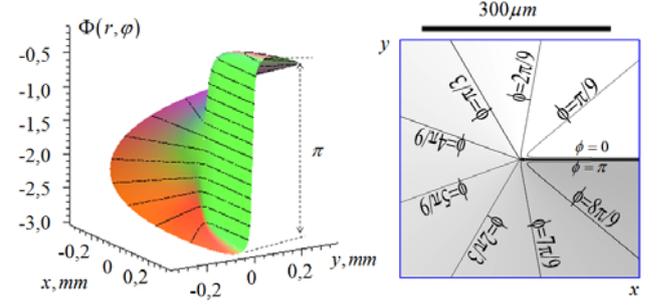

Fig.2 Phase diagram of the first type erf-G beam near the axis with $w_0 = 150 \mu m$ at z=0

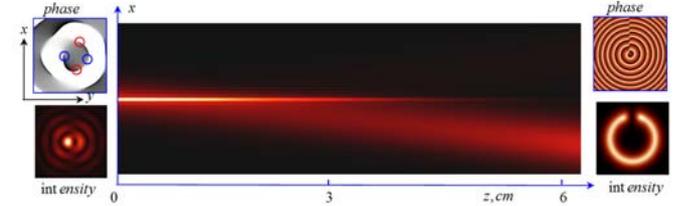

Fig.3. (Color online) Sections *x-y* and *x-z* of the phase and intensity in the erf-G beam of the second

**2.** It is convenient to describe the vector potential for the TE and TM vortex-beams in the basis of circular polarizations [12]: $A_+ = A_x - iA_y$, $A_- = A_x + iA_y$ in the variables $u = x + iy$, $v = x - iy$. In the paraxial approximation one also has the additional variables [11] $V = X - iY$, $U = X + iY$.

In the case of the TE mode beam ($E_z = 0$, $A_z = 0$) we have from eq. (1):

$\partial_V A_+ + \partial_U A_- = 0$ and $A_+ = \partial_U \Psi / Z$, $A_- = -\partial_V \Psi / Z$, (9)

where we made use of the commutation property:

$$\left[ \hat{D}, \partial_U^{(m)} \right] = 0, \quad \left[ \hat{D}, \partial_V^{(n)} \right] = 0, \quad (10)$$

while $\hat{D} \equiv \partial^2 / \partial X^2 + \partial^2 / \partial Y^2$ is the operator of eq. (3), $m, n = 1, 2, 3, \dots$.

Thus, the electric and magnetic fields of the TE mode beam are

$E_+ = ik\partial_U \Psi / Z$, $E_- = -ik\partial_V \Psi / Z$, $E_z = 0$, (11)
$H_+ \approx k\partial_U \Psi / Z$, $H_- \approx k\partial_V \Psi / Z$, $H_z \approx i\partial_{UV}^2 \Psi / Z^2$. (12)

In the case of the TM mode beam we obtain

$A_+ = \partial_U \Psi / Z$, $A_- = \partial_V \Psi / Z$, $A_z = 0$, (13)

so that the electric and magnetic fields take the form

$E_+ = ik\partial_U \Psi / Z$, $E_- = ik\partial_U \Psi / Z$, $E_z \approx \partial_{UV}^2 \Psi / Z^2$, (14)
$H_+ \approx k\partial_U \Psi / Z$, $H_- \approx -k\partial_V \Psi / Z$, $H_z = 0$. (15)

It is worth to note that the functions:

$$F_U^{(m)} = \partial_U^{(m)} F = (iK)^m \times$$
$$\times \int_0^{2\pi} \exp\left\{ i\left[ (p-m)\varphi' + KR\cos(\varphi' - \varphi) \right] \right\} d\varphi', \quad (16)$$

$$F_V^{(n)} = \partial_V^{(n)} F = (iK)^n \times$$
$$\times \int_0^{2\pi} \exp\left\{ i\left[ (p+n)\varphi' + KR\cos(\varphi' - \varphi) \right] \right\} d\varphi', \quad (17)$$

are the solutions of eq. (3) and can serve as the basis for constructing a new set of TE and TM beams $(m = n)$.

We will consider the simplest case when $m = n = 1$ and focus our attention on the erf-G beams of the second type: $K$ is a pure imaginary value. Fig.4 illustrates evolution of the space variant polarization in such a beam with $K = i10^4$ and $w_0 = 150 \mu m$. At the initial plane z=0 the position of zero intensity point is displaced by the distance $\Delta y \approx 0.1 mm$ along the y-axis relative to the beam axis. Near the axis the polarization distribution is far from the standard one for the TE and TM modes. Nevertheless, the longitudinal components of the electric and magnetic fields vanish in the upper and lower rows of the pictures.

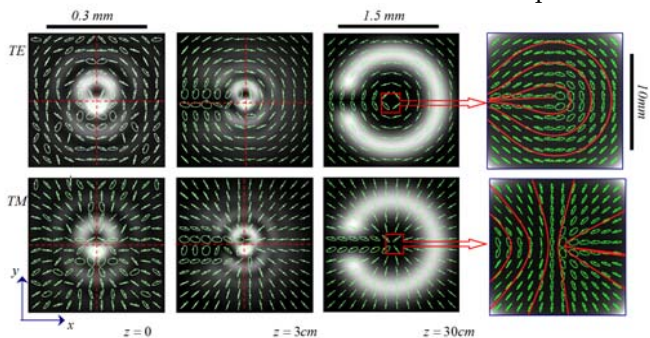

Fig.4. (Color online) Structural transformation of the electric field in asymmetric TE and TM beams with $K = i10^4$ and $w_0 = 150 \mu m$ against a background of the total beam intensity

When spreading the beam the polarization distribution radically changes. The local polarization states tend to the linear ones. Off-axis optical vortices in both circularly polarized components disappear with the exception of the centered singly-charged optical vortex in the RHP component. At the far diffraction field $z >> z_0$ the space variant polarization reproduces nearly the standard form save for the vicinity of the $\varphi = \pi$ ray where intensity vanishes.

The authors are thankful to A. P. Kiselev and E. G. Abramochkin for useful discussion. The paper was partially supported by the grant 0111U008256 of the State fund for fundamental researches of Ukraine.